\documentclass[twocolumn,runningheads,natbib]{svjour2}
\smartqed  
\usepackage{amsfonts,amsmath,amssymb}
\newcommand{\be}{\begin{equation}}
\newcommand{\ee}{\end{equation}}
\newcommand{\bdm}{\begin{displaymath}}
\newcommand{\edm}{\end{displaymath}}
\newcommand{\sun}{\hbox{$\odot$}}
\newcommand{\dmfc}{\dot{\mathfrak{M}_{\rm c}}}
\newcommand{\dmf}{\dot{\mathfrak{M}}}
\journalname{Astrophysics and Space Science}
\begin{document}

\title{Accretion by Isolated Neutron Stars}

\author{N.R.\,Ikhsanov}

\institute{N.R.\,Ikhsanov \at Institute of Astronomy, University of Cambridge,
Madingley Road, Cambridge CB3\,0HA, UK\\
\email{ikhsanov@ast.cam.ac.uk}}

\date{Received: date / Accepted: date}

\maketitle

\begin{abstract}
Accretion of interstellar material by an isolated neutron star is discussed. The point I
address here is the interaction between the accretion flow and the stellar
magnetosphere. I show that the interchange instabilities of the magnetospheric boundary
under the conditions of interest are basically suppressed. The entry of the material
into the magnetosphere is governed by diffusion. Due to this reason the persistent
accretion luminosity of isolated neutron stars is limited to $< 4 \times 10^{26}\,{\rm
erg\,s^{-1}}$. These objects can also appear as X-ray bursters with the burst durations
of $\sim 30$\,minutes and repetition time of $\sim 10^5$\,yr. This indicates that the
number of the accreting isolated neutron stars which could be observed with recent and
modern X-ray missions is a few orders of magnitude smaller than that previously
estimated.
 \keywords{accretion \and neutron stars \and magnetic field}
 \PACS{97.10.Gz \and 97.60.Jd \and 98.38.Am}
\end{abstract}

 \section{Introduction}\label{intro}

As a neutron star moves through the interstellar medium it interacts in a time unit with
the mass
 \be
\dmf_0 \simeq 10^9\,{\rm g\,s^{-1}}\ n\ m^2\ \left(\frac{V_{\rm rel}}{10^7\,{\rm
cm\,s^{-1}}}\right)^{-3},
 \ee
where $m$ is the mass of the neutron star expressed in units of $1.4\,M_{\sun}$, $n$ is
the number density of material situated beyond the accretion (Bondi) radius of the star
expressed in units of 1 hydrogen atom cm$^{-3}$ and $V_{\rm rel}$ is the relative
velocity between the star and its environment, which is limited to the sound speed in
the interstellar material as $V_{\rm rel} > V_{\rm s0}$. The mass capture rate by the
star from its environment is therefore limited to $\dmfc \leq \dmf_0$.

A necessary condition for the captured material to reach the stellar surface is
 \be\label{main}
r_{\rm m} < r_{\rm cor},
 \ee
where
  \be
r_{\rm m} = \left(\frac{\mu^2}{\dmfc \sqrt{2 GM_{\rm ns}}}\right)^{2/7},
 \ee
is the magnetospheric radius of a neutron star, and
 \be
r_{\rm cor} = \left(\frac{GM_{\rm ns} P_{\rm s}^2}{4 \pi^2}\right)^{1/3}.
 \ee
is its corotational radius. Here $\mu$, $M_{\rm ns}$ and $P_{\rm s}$ are the dipole
magnetic moment, mass and spin period of the star, and $G$ is the gravitational
constant. Solving the inequality~(\ref{main}) for $P_{\rm s}$ one finds
 \be
P_{\rm s} > P_{\rm cd} \simeq 7000\ {\rm s}\ \times\ \mu_{30}^{6/7} V_7^{9/7} n^{-3/7}
m^{-11/7},
 \ee
where $\mu_{30} = \mu/10^{30}\,{\rm G\,cm^3}$. This implies that the spin-down rate of
the neutron star in a previous epoch was
 \be
\dot{P} > 10^{-14}\ \left[\frac{P_{\rm s}}{7000\,{\rm s}}\right] \left[\frac{t_{\rm
sd}}{10^{10}\,{\rm yr}}\right]^{-1}\ {\rm s\,s^{-1}},
   \ee
and therefore, suggests that only the stars whose initial dipole magnetic moment was in
excess of $10^{29}\,{\rm G\,cm^3}$ could be a subject of further consideration
\citep[for a discussion see, e.g.,][]{Popov-etal-2000a}. Here $t_{\rm sd}$ is the
spin-down timescale of the neutron star.

Finally, a formation of an accretion disk around the magnetosphere of an isolated
neutron star accreting material from the interstellar medium could be expected only if
the relative velocity satisfies the condition $V_{\rm rel} < V_0$, where
  \be
V_0 \simeq 10^5\ {\rm cm\,s^{-1}}\ \mu_{30}^{-6/65} n^{3/65} m^{5/13}\ \times
 \ee
 \bdm
\left(\frac{V_{\rm t}}{10^6\,{\rm cm\,s^{-1}}}\right)^{21/65} \left(\frac{R_{\rm
t}}{10^{20}\,{\rm cm}}\right)^{-7/65}.
 \edm
Here $V_{\rm t}$ is the velocity of turbulent motions of the interstellar material at a
scale of $R_{\rm t}$ and the Kolmogorov spectrum of the turbulent motions is assumed
\citep{Prokhorov-etal-2002}. This inequality, however, can unlikely be satisfied since
$V_0$ is smaller than the speed of sound in the interstellar material, and therefore, is
smaller than the lower limit to $V_{\rm rel}$.

Thus, the accretion by old isolated neutron stars can be treated in terms of a spherical
(Bondi) accretion onto a magnetized, slowly rotating neutron star. The accretion picture
under these conditions has been reconstructed first by \cite{Arons-Lea-1976} and
\cite{Elsner-Lamb-1976} and further developed by \cite{Lamb-etal-1977} and
\cite{Elsner-Lamb-1984}. An application of the results reported in these papers to the
case of an isolated neutron star is discussed in the following sections.

   \section{Accretion flow at $r_{\rm m}$}

As shown by \cite{Arons-Lea-1976} and \cite{Elsner-Lamb-1976}, the magnetosphere of a
neutron star undergoing spherically symmetrical accretion is closed and, in the first
approximation, prevents the accretion flow from reaching the stellar surface. The mass
accretion rate onto the stellar surface is therefore limited to the rate of plasma entry
into the magnetosphere. The fastest modes by which the material stored over the
magnetospheric boundary can enter the stellar magnetic field are the Bohm diffusion and
interchange instabilities \citep{Elsner-Lamb-1984}.

The rate of plasma diffusion in the considered case can be evaluated as
\citep{Ikhsanov-2003}
 \be
\dmf_{\rm B} \leq 2 \times 10^6\,{\rm g\,s^{-1}}\ \zeta_{0.1}^{1/2}\ \mu_{30}^{-1/14}
m^{15/7} n^{11/14} V_7^{33/14},
  \ee
where $\zeta_{0.1} = \zeta/0.1$ is the efficiency of the diffusion process normalized
according to \cite{Gosling-etal-1991}. This indicates that the luminosity of the
diffusion-driven source is limited to
 \be\label{lxdd}
L_{\rm x,dd} \leq 4 \times 10^{26}\,{\rm erg\,s^{-1}}\ \times
 \ee
 \bdm
 \zeta_{0.1}^{1/2}\
\mu_{30}^{-1/14}\ m^{22/7}\ n^{11/14}\ V_7^{33/14}\ r_6^{-1}.
 \edm

For the material to enter the stellar magnetic field with the rate $\sim \dmfc$ the
boundary should be interchange unstable. The onset condition for the instabilities is
\citep{Arons-Lea-1976,Elsner-Lamb-1976}
 \be\label{inst}
T_{\rm p}(r_{\rm m}) \leq 0.1 T_{\rm ff}(r_{\rm m}),
 \ee
where $T_{\rm p}(r_{\rm m})$ and $T_{\rm ff}(r_{\rm m})$ are the plasma temperature and
the free-fall (adiabatic) temperature at the magnetospheric boundary, respectively. This
indicates that a direct accretion of the captured material onto the stellar surface
could occur only if the cooling of the plasma at the boundary dominates the heating.

The mechanism which is responsible for the cooling of plasma at the boundary is the
bremsstrahlung emission. Indeed, the magnetospheric radius, free-fall temperature and
number density of the material stored over the boundary are, respectively,
 \be
r_{\rm m} \simeq 6\times 10^{10}\,{\rm cm}\ \times\ \mu_{30}^{4/7} \dmf_9^{-2/7}
m^{-1/7},
 \ee
 \be
T_{\rm ff}(r_{\rm m}) \simeq\ 10^7\,{\rm K}\ \times\ \mu_{30}^{-4/7} \dmf_9^{2/7}
m^{6/7},
 \ee
 \be
N_{\rm e}(r_{\rm m}) \simeq 300\,{\rm cm^{-3}}\ \times\ \mu_{30}^{-6/7} \dmf_9^{10/7}
m^{-2/7}.
 \ee
Under these conditions both the cyclotron and Compton cooling time scale are
significantly larger than the bremsstrahlung cooling time scale
 \be
t_{\rm br}(r_{\rm m}) \simeq 10^5\,{\rm yr}\ \times\ T_7^{1/2} \left(\frac{N_{\rm
e}(r_{\rm m})}{300\,{\rm cm^{-3}}}\right)^{-1},
 \ee
where $T_7 = T_{\rm ff}(r_{\rm m})/10^7$\,K.

The heating of the material at the magnetospheric boundary is governed by the following
processes.

  \subsection{Adiabatic shock}\label{shock}

As the captured material reaches the boundary it stops in an adiabatic shock. The
temperature in the shock increases to $T_{\rm ff}(r_{\rm m})$ on a dynamical time scale,
 \be
t_{\rm ff}(r_{\rm m}) \simeq 740\,{\rm s}\ \times\ m^{-1/2} \left(\frac{r_{\rm
m}}{6\times 10^{10}\,{\rm cm}}\right)^{3/2}.
 \ee
Since $t_{\rm ff}(r_{\rm m}) \ll t_{\rm br}(r_{\rm m})$ the height of the homogeneous
atmosphere at the boundary proves to be $\sim r_{\rm m}$. This prevents an accumulation
of material over the boundary. Furthermore, as the condition $t_{\rm ff}(r_{\rm G}) <
t_{\rm br}(r_{\rm m})$ is satisfied throughout the gravitational radius of the neutron
star a hot quasi-stationary envelope extended from $r_{\rm m}$ up to $r_{\rm G}$ forms
\citep{Davies-Pringle-1981}. The formation of the envelope prevents the surrounding
material from penetrating to within the gravitational radius of the neutron star. The
mass of the envelope is, therefore, conserved. As the neutron star moves through the
interstellar medium the surrounding material overflow the outer edge of the envelope
with a rate $\dmfc$.

Within an approximation of a non-rotating star whose ``magnetic gate'' at the boundary
is closed completely the envelope remains in a stationary state on a time scale of
$t_{\rm br}(r_{\rm m})$. As the condition~(\ref{inst}) is satisfied the boundary becomes
unstable and material enters into the magnetic field and accretes onto the stellar
surface with a rate of $\sim \dmfc$. As shown by \cite{Lamb-etal-1977}, the time of the
accretion event in this case is limited to $t_{\rm burst} <$\,a few\,$\times t_{\rm
ff}(r_{\rm m})$ during which the temperature of the envelope increases again to the
adiabatic temperature (as the upper layers of the envelope comes to $r_{\rm m}$). The
corresponding source, therefore, would appear as an X-ray burster with the luminosity
 \be
L_{\rm burst} \simeq 2 \times 10^{29}\ n\ V_7^{-3}\ m^3\ r_6^{-1}\ {\rm erg\,s^{-1}},
 \ee
the typical outburst durations of $t_{\rm burst} \leq 30$\,min and the repetition time
of $t_{\rm rep} \sim t_{\rm br}(r_{\rm m}) \sim 10^5$\,yr.

  \subsection{Subsonic propeller}

As shown by \cite{Davies-Pringle-1981}, the rotation of a neutron star surrounded by the
hot envelope can be neglected only if its spin period exceeds \citep{Ikhsanov-2001}
 \be
P_{\rm br} \simeq 10^5\,{\rm s}\ \times\ \mu_{30}^{16/21} n^{-5/7} V_7^{15/7}
m^{-34/21}.
 \ee
Otherwise, the heating of plasma at the inner edge of the envelope due to propeller
action by the star dominates cooling. The corresponding state of the neutron star is
referred to as a subsonic propeller. The star remains in this state as long as its spin
period satisfies the condition $P_{\rm cd} < P_{\rm s} < P_{\rm br}$. The time during
which the spin period increases from $P_{\rm cd}$ up to $P_{\rm br}$ is
 \be
\tau_{\rm br} \simeq 2 \times 10^5\,{\rm yr}\ \times\ \mu_{30}^{-2}\ I_{45}\ m\
\left(\frac{P_{\rm br}}{10^5\,{\rm yr}}\right),
 \ee
where $I_{45}$ is the moment of inertia of the neutron star expressed in units of
$10^{45}\,{\rm g\,cm^2}$. This indicates that the spin periods of accreting isolated
neutron stars are expected to be in excess of a day, and therefore, these objects can
unlikely be recognized as pulsars.

  \subsection{Diffusion-driven accretor}

As mentioned above, the ``magnetic gate'' at the magnetospheric boundary is not closed
completely. The plasma flow through the interchange stable boundary is governed by the
diffusion. As shown by \cite{Ikhsanov-2003}, this leads to a drift of the envelope
material towards the star and, as a result, to an additional energy source for heating
of the envelope material. The heating due to the radial drift dominates the
bremsstrahlung energy losses from the envelope if
 \be
\dmfc \leq 10^{14}\,{\rm g\,s^{-1}}\ \times\ \zeta_{0.1}^{7/17} \mu_{30}^{-1/17}
V_{7}^{14/17} m^{16/17}.
 \ee
This indicates that only the old isolated neutron stars which move slowly ($V_{\rm rel}
\ll 10^7\,{\rm cm\,s^{-1}}$) through a dense molecular cloud ($N_{\rm e} > 10\,{\rm
cm^{-3}}$) can be expected to be observed as the bursters. The rest of the population
would appear as persistent X-ray sources with the luminosity of $L_{\rm x} \leq L_{\rm
x,dd}$ (see Eq.~\ref{lxdd}).

  \section{Discussion}

The results of this paper force us to reconsider previously made predictions about the
number of old isolated neutron stars which could be observed with recent and current
X-ray missions. In particular, the total flux of the persistent emission of these
objects within the above presented accretion scenario is limited to $F < 10^{-16}\,{\rm
erg\,cm^2\,s^{-1}}\,d_{100}^{-2}$, where $d_{100}$ is the distance to the source
expressed in units of 100\,pc. Furthermore, the mean energy of photons emitted by these
objects within the blackbody approximation is close to 50\,eV. This clearly shows that a
detection of these sources by the Chandra and XMM-Newton is impossible.

The X-ray flux emitted by the accreting isolated neutron stars during the outbursts (see
Sect.\,\ref{shock}) is over the threshold of sensitivity of modern detectors. However,
the probability to detect this event appears to be negligibly small. Indeed, the number
of these sources which could be detected by Chandra and XMM-Newton is
 \be
N \leq 10^{-5} \left(\frac{N(0)}{3 \times 10^4}\right) \left(\frac{t_{\rm
burst}}{30\,{\rm min}}\right) \left(\frac{t_{\rm rep}}{10^5\,{\rm yr}}\right),
 \ee
where $N(0)$ is the number of the sources which would be observed if the influence of
the stellar magnetic field to the accretion flow at $r_{\rm m}$ were insignificant
\citep{Popov-etal-2000b}.

Our results, therefore, naturally explain a lack of success in searching for the
isolated neutron stars accreting material from the interstellar medium. They rather
suggest that these objects can be considered as targets for coming space missions with
more sensitive detectors in the soft X-ray part of the spectrum.

\begin{acknowledgements}
I acknowledge the support of the European Commission under the Marie Curie Incoming
Fellowship Program.
\end{acknowledgements}

\end{document}